\documentclass[eqsecnum,showpacs,twocolumn,floatfix,superscriptaddress]{revtex4}
\usepackage{graphicx}
\usepackage{amsmath}
\begin{document}
\title{Feedback of the electromagnetic environment on current and voltage  fluctuations out of equilibrium}
\author{M. Kindermann}
\affiliation{Instituut-Lorentz, Universiteit Leiden, P.O. Box 9506, 2300 RA
Leiden, The Netherlands}
\author{Yu.\ V. Nazarov}
\affiliation{Department of Nanoscience, Delft University of Technology,
Lorentzweg 1, 2628 CJ Delft, The Netherlands}
\author{C. W. J. Beenakker}
\affiliation{Instituut-Lorentz, Universiteit Leiden, P.O. Box 9506, 2300 RA
Leiden, The Netherlands}
\date{June 2003}

\begin{abstract}
A  theory is presented for low-frequency current and
voltage correlators of a mesoscopic conductor embedded in a
macroscopic electromagnetic environment. This Keldysh field theory evaluated at its saddle-point
provides the microscopic justification for our earlier
phenomenological calculation (using the cascaded Langevin approach).
The nonlinear feedback from the environment mixes correlators of
different orders, which explains the unexpected temperature dependence
of the third moment of tunneling noise observed in a recent
experiment. At non-zero temperature, current and voltage correlators
of order three and higher are no longer linearly related. We show that
a Hall bar measures voltage correlators in the longitudinal voltage
and current correlators in the Hall voltage. We go beyond the saddle-point  approximation to consider the
environmental Coulomb blockade. We derive that the leading order
Coulomb blockade correction to the $n$-th cumulant of current
fluctuations is proportional to the voltage derivative of the
$(n+1)$-th cumulant, generalizing to any $n$ the earlier results for
$n=1,2$.
\end{abstract}

\maketitle

\section{introduction}
A mesoscopic conductor is part of a macroscopic electrical circuit that influences its transport properties. This electromagnetic environment is a source of decoherence  and plays a central role for single-electron effects \cite{IngNaz,Dev90,Ben83,Sch85,Lee96}. Most studies address time-averaged properties. Time-dependent fluctuations of the electrical current are also affected by the environment, which reduces the low-frequency fluctuations by a feedback loop: A current fluctuation $\delta I$ induces a counter-acting voltage fluctuation $\delta V=-Z\delta I$ over the conductor, which in turn reduces the current by an amount $-G\delta V$. (Here $G$ and $Z$ are, respectively, the conductance of the mesoscopic system and the equivalent series impedance of the macroscopic voltage-biased circuit.) 

At zero temperature the macroscopic circuit does not generate any noise itself, and the feedback loop is the only way it affects the current fluctuations in the mesoscopic conductor, which persist at zero temperature because of the shot noise effect \cite{Shulman,Bla00,Bee02}. In the second cumulant $C^{(2)}$, or shot-noise power, the feedback loop may be accounted for by a rescaling of the current fluctuations: $\delta I\rightarrow (1+ZG)^{-1}\delta I$. For example, the Poisson noise $C^{(2)}=e\bar{I}(1+ZG)^{-2}$ of a tunnel junction is simply reduced by a factor $(1+ZG)^{-2}$ due to the negative feedback of the series impedance. We have recently discovered that this textbook result breaks down beyond the second cumulant  \cite{Kin02}. Terms appear which depend in a nonlinear way on lower cumulants, and which can not be incorporated by any rescaling with powers of $1+ZG$. In the example of a tunnel junction the third cumulant at zero temperature takes the form $C^{(3)}=e^{2}\bar{I}(1-2ZG)(1+ZG)^{-4}$.

Ref.\ \cite{Kin02} was restricted to zero temperature. In Ref.
 \cite{Bee03} we removed this restriction  and showed that the nonlinear feedback of the electromagnetic environment drastically modifies the temperature dependence of $C^{(3)}$.  Earlier theory   \cite{Lev01,Gut02,Nag02} assumed an isolated mesoscopic conductor and predicted a  temperature-independent $C^{(3)}$ for a tunnel junction. The coupling to an environment introduces a temperature dependence, which can even  change the sign of $C^{(3)}$ as the temperature is raised.   No such effect exists for the second cumulant.
The predicted temperature dependence has been measured in a recent experiment \cite{Reu03}.  The method we used in Ref.\ \cite{Bee03} to arrive at these results was phenomenological. The nonlinear feedback was inserted by hand into the Langevin equation, through a cascade assumption \cite{Nag02}. The purpose of the present paper  is to provide a fully quantum mechanical  derivation. Our results  agree with Ref.\ \cite{Bee03}, thereby justifying    the Langevin approach.

The outline of this paper is as follows. In Secs.\ \ref{sc:formulation} and \ref{sc:calculation} we  present the general  framework within which we  describe a broad class of  electrical circuits that consist of mesoscopic conductors embedded in a macroscopic   electromagnetic environment.   The basis is a path integral formulation of the Keldysh approach to charge counting statistics  \cite{Naz99,Kin01}. It allows us to compute correlators and cross-correlators of currents and voltages at arbitrary contacts of the circuit.  The method is technically involved, but we give an intuitive interpretation of the results in terms of  ``pseudo-probabilities''.  Within this framework we study in Secs.\ \ref{sc:linres} and \ref{sc:appl}  series circuits of two conductors.   For concrete results we specialize to a low-frequency regime where the path integrals over fluctuating quantum  fields can be taken in saddle-point approximation. The conditions of validity for this  approximation are discussed.  We obtain general relations between third order correlators in a series circuit and   correlators of the individual isolated conductors. We specialize to the experimentally relevant case of a single mesoscopic conductor in series with a macroscopic conductor that represents  the    electromagnetic environment.   Most experiments measure voltage correlators. In Sec.\ \ref{sc:current} we propose an experimental method to obtain current correlators, using the Hall voltage in a weak magnetic field.  The fundamental difference between current and voltage correlators rests on whether the variable measured is odd or even under time reversal.  In Sec.\ \ref{sc:Coulomb} we relax the low-frequency approximation  by  addressing Coulomb blockade effects from the environment \cite{Yey01,Gal02,Kin03}.  We conclude in Sec.\ \ref{sc:conclusion}.

\section{Description of the circuit} \label{sc:formulation}

\begin{figure}
\includegraphics[width=8cm]{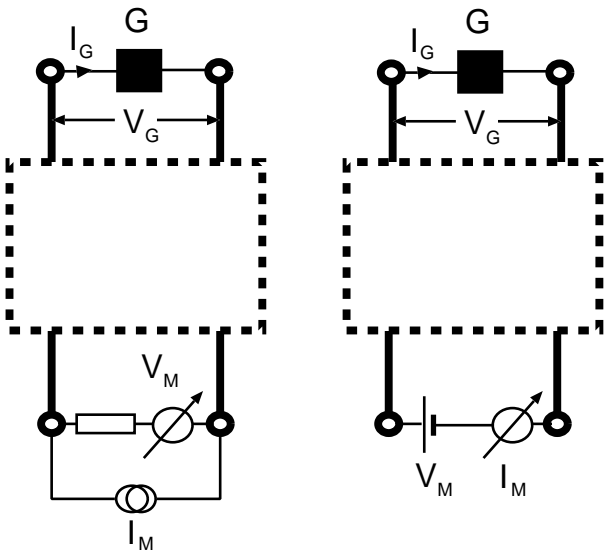}
\caption{Electrical circuits  studied in this article. The black boxes represent  conductors embedded in an electromagnetic environment (dashed rectangle). A voltage source is present at the contacts for a current measurement (right circuit) and a current source at the contacts for a  voltage measurement (left circuit). The two circuits can also be combined into one larger circuit containing two conductors and both a current and a voltage meter. }
\label{multiterminal}
\end{figure}
We consider a circuit consisting of electrical conductors $G_i$, a macroscopic electromagnetic environment [with impedance matrix $Z(\omega)$], plus ideal current and voltage meters $M_i$. The current meter (zero internal impedance) is in series with a voltage source, while the voltage meter (infinite internal impedance) is in parallel to a current source. Any finite impedance of meters and sources is incorporated in the electromagnetic environment. In Fig.\ \ref{multiterminal} we show examples of such circuits.

\begin{figure}
\includegraphics[width=3.6cm]{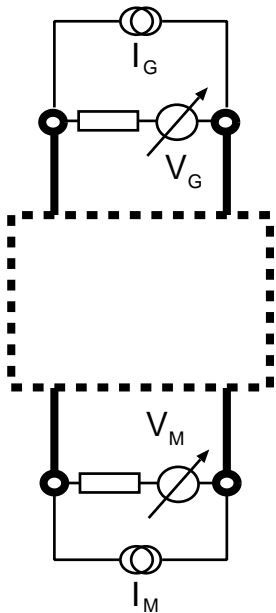}
\caption{
Circuit used to characterize the impedance matrix of the electromagnetic environment. All contacts are now connected to a voltage meter  plus a current source.  \label{linear}
 }
\end{figure}

 The electromagnetic environment is assumed to produce only thermal noise.
  To characterize this noise we consider the circuit without the mesoscopic conductors, see Fig.\ \ref{linear}. Each pair of contacts to the environment  is now attached to a current source and a voltage meter. The impedance matrix is defined by partial derivatives of voltages with respect to currents,
\begin{equation} \label{eq:Zmatrix}
\boldsymbol{Z} =   \left( \begin{array}{cc}
{Z_{GG}} &  {Z_{GM}} \\
{Z_{MG}}  & {Z_{MM}}  \end{array} \right) =    \left( \begin{array}{cc}
\frac{\partial V_G}{\partial I_G}\Big|_{I_M} &  \frac{\partial V_G}{\partial I_M}\Big|_{I_G}  \\
\frac{\partial V_M}{\partial I_G}\Big|_{I_M}   & \frac{\partial V_M}{\partial I_M}\Big|_{I_G}   \end{array} \right).
\end{equation}
(All quantities are taken at the same frequency $\omega$.)
 If there is more than one pair of contacts of type $G$ or $M$, then the four blocks of $\boldsymbol{Z}$ are themselves matrices. Positive and negative frequencies are related by $Z_{\alpha \beta}(-\omega)=Z_{\alpha \beta}^*(\omega)$. We also note the Onsager-Casimir \cite{Cas45} symmetry $Z_{\alpha\beta}(B,\omega)=Z_{\beta \alpha}(-B,\omega)$, in an external magnetic field $B$.   The thermal noise at each pair of contacts is Gaussian. The covariance matrix of the voltage fluctuations $\delta V_{\alpha}$ is  determined by the fluctuation-dissipation theorem,
 \begin{eqnarray} \label{eq:Gaussnoise}
  \langle \delta V_{\alpha} (\omega)  \delta V_{\beta}  (\omega') \rangle &=&  \pi  \delta(\omega+\omega') \hbar \omega \coth\left(\frac{\hbar \omega}{2k T}\right) \nonumber \\
  && \mbox{} \times [Z_{\alpha \beta}(\omega)+Z^*_{\beta\alpha}(\omega)], 
  \end{eqnarray}
 with  $T$ the temperature of the  environment.

We seek finite frequency cumulant correlators of the  variables measured at the current and voltage meters,
\begin{equation} \label{eq:corrdef}
\langle\!\langle X_{1}(\omega_{1})\cdots  X_{n}(\omega_{n})\rangle\!\rangle=2 \pi \delta \left(\sum_{k=1}^n \omega_{k} \right){ C}_{\bf X}^{(n)}(\omega_{1},\cdots,\omega_n).
\end{equation}
Here $X_i$ stands for either $V_M$ or $I_M$.  Fourier transforms are defined by  $X_{i}(\omega) = \int{dt\, \exp(i \omega t) X_{i}(t)}$. 
Our aim is to relate the  correlators  at the measurement contacts to the  correlators one would measure at the conductors if they were   isolated from the environment.

 \section{Path integral formulation} \label{sc:calculation}

Correlators of   currents $I_M$ and  voltages $V_M$ at  the measurement contacts are obtained from the generating functional
\begin{equation} \label{eq:Zvolt}
 {\cal Z}_{\boldsymbol{ X}} [ \boldsymbol{ \vec{j}_{ }}]= \left\langle  
 { T}_- e^{i \int{dt\; \left[{H} +\boldsymbol{  j^-}(t)  \boldsymbol{ X}\right]} }  { T}_+ e^{-i \int{dt\; \left[{H} + \boldsymbol{ j^+}(t)  \boldsymbol{ X}\right]} } \right\rangle.
\end{equation}
They contain moments of outcomes of measurements of the variable $X$ (equal to $I_M$ or $V_M$) at different instants   of time. The symbols ${T}_+({T}_-)$ denote  (inverse)
 time ordering, different on the forward and the backward part of the Keldysh contour.
 The exponents contain source terms $j^{\pm}$ and a Hamiltonian $H$, which  we discuss separately. 
 
 The source term $j^{\pm}(t)$  is a charge $Q_M=\int^t{dt' \, I_M(t')}$ if $X=V_M$, whereas  it is a phase $\Phi_{{ M}}=\int^t{dt' \, V_M(t')}$ if  $X=I_{{ M}}$. (We have set $\hbar$ to unity.)
  The superscript $\pm$ determines  on which part of the Keldysh  contour the source  is effective. The vector $\boldsymbol{ \vec{j}}= (\boldsymbol{j^{cl}},\boldsymbol{j^q})$   indicates the linear combinations      \begin{equation}
  \boldsymbol{ j^{cl}}=\frac{1}{2}\frac{\partial}{\partial t}(\boldsymbol{ j^+} +\boldsymbol{ j^-}), \;\;\; \boldsymbol{ j^{q}}=\boldsymbol{ j^+} -\boldsymbol{ j^-}.
  \end{equation}
     We denote vectors in this two-dimensional ``Keldysh space'' by a vector arrow.   The 
``classical'' source fields $\boldsymbol{ {j}^{cl}_{{ }}}=(j_1^{cl},j_2^{cl},\cdots)$  account for   current or voltage sources at the measurement  contacts.  Cumulant correlators of the measured variables are generated by differentiation of $\ln {\cal Z}_{\boldsymbol{ X}}$  with respect to the ``quantum'' fields  $\boldsymbol{ {j}^{q}_{{ }}}=(j_1^{q},j_2^{q},\cdots)$: 
\begin{equation} \label{eq:funcdiff}
\left\langle\! \left\langle \prod_{k=1}^n X_{k}(t_{k})  \right\rangle \! \right\rangle = \prod_{k=1}^n \frac{\delta}{ -i \delta j^q_{k}(t_{k})}  \ln { {\cal Z}}_{\boldsymbol{ X}} \Big|_{\boldsymbol{ {j}^q }=0}.
\end{equation}
 
 The   Hamiltonian consists of three parts,
\begin{equation} \label{eq:HamV}
H = H_{ e} + \sum_{i} H_{{G}_{i}} - \boldsymbol{\Phi_{ G}}  \boldsymbol{ I_{ G}}.
\end{equation}
 The term  $H_{ e} = \sum_{j} \Omega_{j} a_j ^{\dagger} a_j$ represents the electromagnetic environment, which we model by a collection of harmonic oscillators at frequencies $\Omega_j$.  The conductors connected to the environment  have Hamiltonians $H_{G_{i}}$. 
 The interaction term couples the phases  $\boldsymbol{\Phi_{ G}}$ (defined by $i[H_e,\boldsymbol{\Phi_G}]=\boldsymbol{V_G}$) to the currents $\boldsymbol{ I_{ G}}$ through the conductors.   The phases $\boldsymbol{\Phi_G}$, as well as the measured quantities ${\bf X}$, are linear combinations of the bosonic operators $a_j$ of the electromagnetic environment,  
 \begin{eqnarray} \label{eq:Henv}
  \boldsymbol{\Phi_{ G}}   &=& \sum_j \left(\boldsymbol{c^{ G}_j} a_j+\boldsymbol{c^{ G*}_j} a^{\dagger}_j\right)  , \\
 \boldsymbol{ X}   &=& \sum_j \left( \boldsymbol{ c^{ X}_j } a_j+  \boldsymbol{ c^{ X*}_j } a^{\dagger}_j\right).
\end{eqnarray}
The coefficients  $c^{ G}_j$ and $c^{ X}_j$ depend on the impedance matrix of the environment and also on which contacts are connected to a current source and which to a voltage source. 

To calculate the generating functional we use a path integral formulation in Keldysh space \cite{Kam01,Kin01}.  We first present the calculation for the case of a voltage measurement at all measurement contacts (so $X_k=V_{M_k}$ and $j_k= Q_{M_k}$ for all $k$).   We will then show how the result for a current measurement can be obtained from this calculation. The path integral involves integrations over the environmental  degrees of freedom $a_{j}$ weighted with an  influence functional    ${\cal Z}_{\boldsymbol{I_{{ G}}}} $
due to the conductors. Because  the  conductors are assumed to be uncoupled in the absence of the environment,  this influence functional factorizes:
\begin{eqnarray} \label{eq:infl}
 {\cal Z}_{\boldsymbol{I_{{ G}}}} [\boldsymbol{\vec{\Phi}_{ G}}] =  \prod_{i}{ {\cal Z}_{I_{{ G}_{i}}} [\vec{\Phi}_{{ G}_i}]}.     
 \end{eqnarray}
 An individual conductor has  influence functional
\begin{eqnarray} \label{eq:inflgamma}
{\cal Z}_{I_{{ G}_i}} &=&  \left\langle  {T}_-\; e^{{i} \int{dt \bigl[ {H}_{{ G}_i}+\Phi_{G_i}^-(t) {I}_{{ G}_i}\bigr] }} \right. \nonumber \\
&&\left.\;\;\;\; \mbox{}\times {T}_+ e^{-{i} \int{dt \bigl[{H}_{{ G}_i}+\Phi_{G_i}^+(t) {I}_{{ G}_i}\bigr]}}
  \right\rangle. 
  \end{eqnarray}
Comparing Eq.\ (\ref{eq:inflgamma}) with Eq.\  (\ref{eq:Zvolt}) for ${\bf X}=\boldsymbol{I_M}$,  we note that the influence functional of a conductor $G_i$ is just the generating functional of current fluctuations in   $G_i$ when connected to an ideal voltage source without  electromagnetic environment. That is why we use the same symbol ${\cal Z}$ for influence functional and generating functional.  

  The integrals over all environmental fields except 
$\boldsymbol{\vec{\Phi}_{ G}}$   are Gaussian  and can be done exactly. 
The resulting path integral  expression for the generating functional ${\cal Z}_{V_M}$ takes the form 
\begin{equation} \label{eq:result}
 {\cal Z}_{\boldsymbol{V_M}}[\boldsymbol{ \vec{Q}_{ M}}]=  \int{ {\cal D}[\boldsymbol{\vec{\Phi}_{ G}}] \exp \left\{-i {\cal S}_{ e}[\boldsymbol{ \vec{Q}_{ M}},\boldsymbol{\vec{\Phi}_{ G}}]   \right\}}\;  {\cal Z}_{\boldsymbol{I_G}}[\boldsymbol{\vec{\Phi}_{ G}}],
\end{equation}
up to a normalization constant \cite{review}.  We use for the integration fields $\boldsymbol{\vec{\Phi}_{ G}}$ the same vector notation as for the source fields:      $\boldsymbol{\vec{\Phi}_{ G}}=(\boldsymbol{ \Phi_{ G}^{cl}},\boldsymbol{\Phi^q_{ G}})$ with  $\boldsymbol{ \Phi_G^{cl}}=\frac{1}{2}(\partial/\partial t)(\boldsymbol{\Phi^+_{ G}} +\boldsymbol{\Phi^-_{ G}})$ and   $\boldsymbol{\Phi_{ G}^q}=\boldsymbol{\Phi^+_{ G}}-\boldsymbol{\Phi^-_{ G}}$.   
The Gaussian environmental action  ${\cal S}_e$ is calculated in App.\  \ref{app:action}. The result is given in terms of the impedance matrix $\boldsymbol{Z}$ of the environment, 
\begin{widetext}
\begin{eqnarray} \label{eq:envaction}
{\cal S}_{ e}[\boldsymbol{ \vec{Q}_{ M}},\boldsymbol{\vec{\Phi}_{ G}}]&=&    {\frac{1}{2}} \int{ \frac{d \omega}{2 \pi} \left[\boldsymbol{ \vec{Q}^{*}_{ M}}\boldsymbol{ \check{Z}_{ MM}} \boldsymbol{ \vec{Q}_{ M}}   \right.} +\left(\boldsymbol{\vec{\Phi}^{*}_{ G}} -  \boldsymbol{ \vec{Q}^{*}_{ M}} \boldsymbol{ \check{Z}_{ MG}} \right) \boldsymbol{\check{Y}} \left(\boldsymbol{\vec{\Phi}_{ G}} - \boldsymbol{ \check{Z}_{ GM}} \boldsymbol{ \vec{Q}_{ M}}\right)\Big],  \\ 
 \label{eq:matrices}
\boldsymbol{ \check{Y}}(\omega) & = & \left( \begin{array}{cc}
0 &   \boldsymbol{ Z}_{ \bf GG}^{\dagger -1 }(\omega) \\
\boldsymbol{ Z}_{ \bf GG}^{-1}(\omega) &   - \frac{i}{2} \omega [2N(\omega)+1]  [\boldsymbol{  Z}_{\bf GG}^{-1}(\omega)+ \boldsymbol{ Z}_{ \bf GG}^{\dagger -1 }(\omega)] 
\end{array} \right),   \\
\boldsymbol{ \check{{Z}}_{ MM}}(\omega)& = & \left( \begin{array}{cc}
0 &  \boldsymbol{ Z}_{\bf MM}^{\dagger}(\omega) \\
  \boldsymbol{ {Z}_{ MM}} (\omega) & -\frac{i}{2}\omega [2N(\omega)+1] [\boldsymbol{ {Z}_{ MM}} (\omega)+ \boldsymbol{ Z}_{\bf  MM}^{\dagger}(\omega)]
\end{array} \right),  \\
\label{eq:matrixoff} \boldsymbol{ \check{Z}_{ MG}}(\omega)&=& \left( \begin{array}{cc}
-\boldsymbol{ Z}_{\bf  GM}^{\dagger}(\omega) &   0 \\
 \frac{ i}{2} \omega  [2N(\omega)+1] [\boldsymbol{ Z_{ MG}}(\omega)+\boldsymbol{ Z}_{\bf GM}^{\dagger}(\omega) ] & \boldsymbol{ Z_{ MG}}(\omega) 
\end{array} \right)= \boldsymbol{ \check{Z}}^T_{\bf GM}(-\omega),
  \end{eqnarray}
  \end{widetext}
   with  the Bose-Einstein distribution $N(\omega)=[\exp( \omega/kT)-1]^{-1}$.    We have marked matrices in the Keldysh space by a check, for instance $\check{\bf Y}$. 
  
   When one substitutes Eq.\ (\ref{eq:envaction}) into Eq.\ (\ref{eq:result})  and calculates correlators with the help of Eq.\ (\ref{eq:funcdiff}), one can identify two sources of noise.
   The first source of noise  is current  fluctuations in the  conductors  that induce  fluctuations of  the measured voltage. These contributions  are generated by differentiating the terms  of ${\cal S}_{ e}$ that are linear in $\boldsymbol{ \vec{Q}_{ M}}$.  The second source of noise is the environment  itself, accounted for by the contributions quadratic in $\boldsymbol{ \vec{Q}_{ M}}$. 

Generating functionals ${\cal Z}_{I_M}$ for circuits  where currents rather than voltages are measured at some of the contacts can  be obtained along the same lines with modified response functions. It is also possible to obtain them from ${\cal Z}_{V_M}$ through the functional Fourier transform derived in App.\ \ref{app:ZIZV}, 
\begin{eqnarray}  \label{eq:ZIZV}
{\cal Z}_{ I_{ M}}[\vec{\Phi}_{{ M}}]= \int{  {\cal D}[\vec{Q}_{{ M}}] \, e^{-i\vec{Q}_{{ M}}\times \vec{\Phi}_{{ M}} }\,
{\cal Z}_{ V_{{ M}}}[\vec{Q}_{{ M}}]}.  
\end{eqnarray} 
We have defined the cross  product
\begin{equation} \label{eq:cross}
 \vec{Q} \times \vec{\Phi} \equiv \int{dt\, (Q^{cl}\Phi^q-\Phi^{cl} Q^q )}.
\end{equation}
  This transformation may be applied to any pair of measurement contacts to obtain current correlators from voltage correlators.

Eq.\ (\ref{eq:ZIZV}) ensures that the two functionals
\begin{eqnarray} \label{eq:pseudo1}
{\cal P}[V,I] &= & \int{ {\cal D}[q] \,e^{i\int{dt \, q V}} \;{\cal Z}_{V} \left[\vec{Q}=(I,q)\right]},\\  
{\cal P}'[V,I] &=& \int{  {\cal D}[\varphi] \,e^{i\int{dt \, \varphi I}}\; {\cal Z}_{I}\left[\vec{\Phi}=(V,\varphi) \right]},  \label{eq:pseudo2} 
\end{eqnarray}
are identical: ${\cal P}[V,I]={\cal P}'[V,I]$.  This functional ${\cal P}$ has an intuitive probabilistic interpretation. 
   With the help of Eq.\ (\ref{eq:funcdiff}) we obtain from ${\cal P}$ the correlators  
\begin{eqnarray} \label{eq:voltint}
 \left\langle  V(t_{1}) \cdots  V(t_{n})  \right\rangle_{I}& =&\frac{\int{{\cal D}[V]\, V(t_{1})\cdots  V(t_{n}) {\cal P}[V,I]}}{\int{{\cal D}[V]\,  {\cal P}[V,I]}}, \nonumber \\
\\ \label{eq:currint}
 \left\langle    I(t_{1}) \cdots I(t_n)  \right\rangle_V &=& \frac{\int{{\cal D}[I]\,   I(t_{1}) \cdots I(t_n)  {\cal P}[V,I]}}{\int{{\cal D}[I]\,  {\cal P}[V,I]} }. \nonumber \\
\end{eqnarray}
This suggests the interpretation of ${\cal P}[V,I]$ as   a joint probability distribution functional of current and voltage fluctuations.   Yet,  ${\cal P}$ can  not properly  be called   a probability since it need not be positive.  In the low frequency approximation introduced in the next section it is positive for  normal metal conductors. However, for  superconductors, it has been found to take negative values \cite{Bel01}. It is therefore more properly called a    ``pseudo-probability''.

   We conclude this section with some remarks on the actual measurement process. 
The time-averaged correlators  (\ref{eq:corrdef}) may be measured in two different ways.  In the first way  the variable $X$ is measured repeatedly and results at different times are correlated afterwards. In the second way (and this is how it is usually done \cite{Gav01}) one uses a detector that measures directly time integrals of  $X$ (for example, by means of a spectral filter).  The correlators measured in the first way are obtained from the generating functional according to Eq.\ (\ref{eq:funcdiff}), 
\begin{eqnarray}
&&2 \pi \delta \left(\sum_{k=1}^n \omega_{k} \right){ C}_{\bf X}^{(n)}(\omega_{1},\cdots,\omega_n) \nonumber \\
&&=
\prod_k \left[ \int_{-\infty}^{\infty}{dt\,e^{i\omega_k t} \frac{\delta}{ -i \delta j^q_{k}(t)}}\right] \ln{ {\cal Z}}_{\boldsymbol{ X}} \Big|_{\boldsymbol{ {j}^q}=0}. 
\end{eqnarray}
The second way of measurement is modelled  by choosing cross-impedances that ensure that an instantaneous measurement at one pair of contacts yields a time average at another pair, for example $Z_{{ MG}} (\omega)\propto \delta(\omega-\omega_0)$. The resulting frequency dependent correlators do not depend on which way of measurement one uses.

\section{Two conductors in series} \label{sc:linres}

\begin{figure}
\includegraphics[width=8cm]{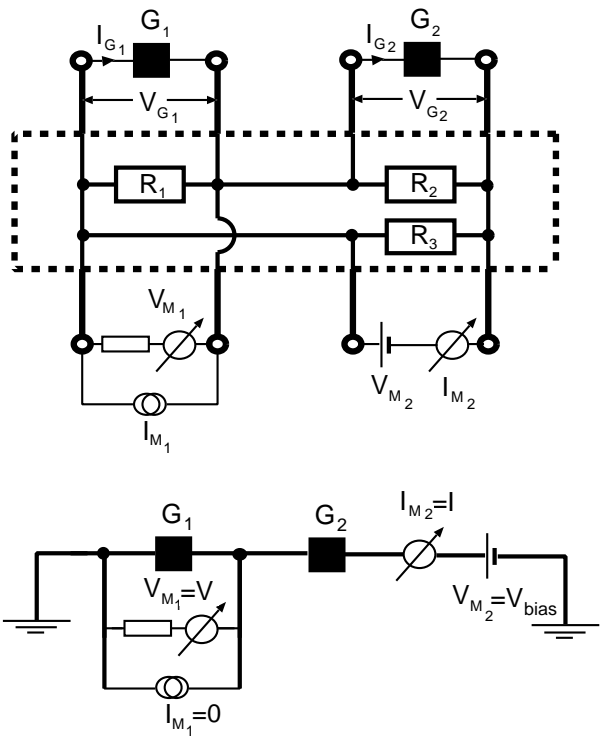}
\caption{Top panel: Circuit   of two conductors $G_1$, $G_2$ in an electromagnetic environment modelled by three resistances $R_1$, $R_2$,  $R_3$. In the limit $R_1, R_2, R_3\to\infty$  the circuit becomes equivalent to the series circuit in the lower panel.}
\label{eightterminal}
\end{figure}

We  specialize the general theory     to the series circuit of  two conductors ${ G}_1$ and ${ G}_2$ shown in Fig.\ \ref{eightterminal} (lower panel). We derive the generating functional  ${\cal Z}_{V,I}$ for correlators of the voltage drop  $V\equiv V_{M_1}$ over  conductor  ${ G}_1$  and the current $I\equiv I_{{ M}_2}$ through both conductors. (The voltage drop over conductor $G_2$ equals $V_{M_2}-V_{M_1}\equiv V_{\rm bias}-V$, with $V_{\rm bias}$ the non-fluctuating bias  voltage of the voltage source.) To apply the general relations of the previous  section we embed the two conductors in an electromagnetic environment, as shown in the top panel of Fig.\ \ref{eightterminal}. In the limit of infinite resistances $R_1$, $R_2$,  and $R_3$  this 8-terminal circuit becomes equivalent to a simple series circuit of $G_1$ and $G_2$.  We take the infinite resistance limit of Eq.\ (\ref{eq:result}) in App.\ \ref{app:8term}. The result 
\begin{equation}  \label{eq:convolution}
{\cal Z}_{V,I}[\vec{Q},\vec{\Phi}]=   \int{  {\cal D}[\vec{\Phi'}] \, e^{-i\vec{\Phi}'\times \vec{Q} }\,
  {\cal Z}_{1} [\vec{\Phi}'] {\cal Z}_{2} [\vec{\Phi}-\vec{\Phi}']} 
\end{equation}
shows that
the generating functional of current and voltage correlators in the series circuit is a  functional integral convolution of  the generating functionals ${\cal Z}_1\equiv {\cal Z}_{I_{G_1}}$ and ${\cal Z}_2\equiv {\cal Z}_{I_{G_2}}$  of the two conductors ${ G}_1$ and ${ G}_2$ defined in Eq.\ (\ref{eq:inflgamma}).

Eq.\ (\ref{eq:convolution}) implies a simple relation between the pseudo-probabilities ${\cal P}_{G_1+G_2}$ of the series circuit (obtained by means of  Eq.\ (\ref{eq:pseudo2}) from ${\cal Z}_{V,I}|_{\vec{Q}=0}$) and the pseudo-probabilities ${\cal P}_{G_k}$  of the individual conductors  (obtained by means of Eq.\ (\ref{eq:pseudo2}) from ${\cal Z}_{k}$). We find 
\begin{equation} \label{eq:pseudorel}
{\cal P}_{ G_1+G_2}[V,I]=
\int{{\cal D}V' \,  
{\cal P}_{ G_1}[V-V',I]
{\cal P}_{ G_2}[V',I]}.
\end{equation}
This relation is obvious  if one interprets it
in terms of classical probabilities:  The voltage drop over $G_1+G_2$ is the sum of the independent voltage drops over $G_1$ and $G_2$, so the probability ${\cal P}_{G_1+G_2}$  is the convolution of ${\cal P}_{G_1}$ and ${\cal P}_{G_2}$. Yet the relation (\ref{eq:pseudorel}) is for quantum mechanical  pseudo-probabilities.

We evaluate the convolution (\ref{eq:convolution}) in the low-frequency regime, when the functionals ${\cal Z}_1$ and ${\cal Z}_2$ become local in time,
\begin{equation}
\ln {\cal Z}_{k}[\vec{\Phi}] \equiv  -i {\cal S}_{k}[\vec{\Phi}]=  -i \int{dt\,S_{ k}\biglb(\vec{\Phi}(t)\bigrb)}.
\end{equation}
   We  then do the path integration  in saddle-point approximation, with the result 
\begin{eqnarray} \label{eq:saddlepointF}
\ln  {\cal Z}_{V,I}[\vec{Q},\vec{\Phi}]&=&  -i \mathop{\rm  extr}\limits_{[\vec{\Phi}']}  \Bigl\{ \vec{\Phi}' \times \vec{Q} \nonumber \\
&&\!\!\!\!\!\!\!\!\!\!\!\!\!\!\!\!\!\! \!\!\!\!\!\!\!\!\!\!\!\!\!\!\!\!\!\! \mbox{}+ \int{dt\, \bigl[{ S}_{ 1}\biglb( \vec{ \Phi}'(t)\bigrb)}  + S_{ 2}\biglb(\vec{\Phi}(t)-\vec{\Phi}'(t)\bigrb)\bigr] \Bigr\}.
\end{eqnarray}
The notation ``extr'' indicates the extremal value of the expression between curly brackets with respect to variations of $\vec{\Phi}'(t)$. 
The validity of the low-frequency and saddle-point approximations is addressed at the end of this section. 

We will consider separately the case that both conductors $G_1$ and $G_2$ are mesoscopic conductors and the case that $G_1$ is mesoscopic while $G_2$ is a macroscopic conductor. The action of a macroscopic conductor with impedance $Z$ is quadratic, 
\begin{eqnarray} \label{eq:envac}
  {\cal S}_{ \rm macro}[\vec{\Phi}] = {\frac{1}{2}} \int{\, \frac{d \omega}{2 \pi}\,  \vec{\Phi}^{\dagger}\check{Y} \vec{\Phi}},
\end{eqnarray}
corresponding to Gaussian current fluctuations. 
The matrix $\check{Y}$ is given by Eq.\ (\ref{eq:matrices}), with a scalar $Z_{ GG}=Z$.   The corresponding pseudo-probability  ${\cal P}_{\rm macro}$ is positive,  
\begin{equation} \label{eq:Penv}
 {\cal P}_{{\rm macro} }[V,I] =  \exp\left\{ - \int{\frac{d\omega}{4\pi\omega} \,\frac{|V-Z I |^2 }{{\rm Re}\,Z}}\tanh \left(\frac{\omega}{2 k T} \right)\right\}.
\end{equation}

Substitution of ${\cal P}_{\rm macro}$ for ${\cal P}_{G_2}$ in Eq.\ (\ref{eq:pseudorel})  gives a simple result for  ${\cal P}_{G_1+G_2}$  at zero temperature,  
\begin{equation} \label{eq:pseudorellin}
{\cal P}_{ G_1+G_2}[V,I]=
{\cal P}_{ G_1}[V-ZI,I], \;\;{\rm if} \;\; T=0.
\end{equation}
The feedback of the macroscopic conductor on the mesoscopic conductor amounts to a negative voltage $-ZI$ produced in response to a current $I$. 
 
The action of a mesoscopic conductor in the low-frequency limit is given by the Levitov-Lesovik formula \cite{Lev93,Lev96},
\begin{eqnarray} \label{eq:Sxi0}
S_{\rm  meso} (\vec{\Phi})&=&  \frac{1}{2\pi} \sum_{n=1}^N \int {d\varepsilon \,\ln
[1+\Gamma_n (e^{i e \varphi}-1) n_{ R} (1-n_{ L})} \nonumber \\
&&\;\;\;\;\;\;\;\;\;\;\;\;\mbox{}+\Gamma_n (e^{-ie  \varphi}-1) n_{ L} (1-n_{ R})], 
\end{eqnarray}
with $\vec{\Phi}=(V,\varphi)$.
The $\Gamma_n$'s ($n=1,2,\cdots N$) are the transmission eigenvalues of the conductor. The two functions  $n_{ L}(\varepsilon,T)= [\exp(\varepsilon/k T)+1]^{-1}$ and $n_{ R}(\varepsilon,T)=n_{ L}(\varepsilon+eV,T)$ are the filling factors of electron states at the left and  right contacts, with $V$ the voltage drop over the conductor and $T$ its temperature.  

The  criterion for the applicability of  the low-frequency and saddle-point approximations to the action of  a mesoscopic conductor  depends on    two time scales.  The first scale  $\tau_1= \min(1/eV,1/kT)$ is the mean width of current pulses due to individual transferred electrons. The second scale  $\tau_{ 2} =e/I\simeq (e^2/ G) \tau_{ 1}$  is the mean time between current pulses.  At frequencies below $1/\tau_1$ the action of the conductor becomes local in time.    Below the second scale $1/\tau_2$ the action of the conductor is large for values of $\vec{\Phi}$ where the nonlinearities become important. This justifies the saddle-point approximation.   The nonlinearities in ${\cal S}_{\rm meso}$ become relevant
for $\varphi\simeq 1/e$, so  for time scales $\tau \gg \tau_2$ we indeed have $   {\cal S}_{\rm meso} \simeq \tau I \varphi  \simeq \tau I/e \,  \simeq \tau/\tau_2 \gg 1$.   

These two approximations together are therefore justified if fluctuations with frequencies $\omega $ above $\Lambda \simeq \min(1/\tau_1,1/\tau_2)$ are suppressed by a small effective impedance:  $Z(\omega)\ll h/e^2$ for $\omega \agt \Lambda$.  A small impedance acts as a heavy mass term  in Eq.\  (\ref{eq:convolution}), suppressing fluctuations. This is seen from  Eq.\ (\ref{eq:envac}) for a macroscopic conductor and it carries over to other conductors.  In Sec.\ \ref{sc:Coulomb} we will examine the Coulomb blockade effects that appear if $Z(\omega)$ is   not small at high frequencies.

\section{Third cumulants} \label{sc:appl}
\subsection{Two arbitrary conductors in series}

We use  the general formula (\ref{eq:saddlepointF}) to calculate the third order cumulant correlator of current and voltage fluctuations  in a series circuit of two conductors $G_1$ and $G_2$ at finite temperature.
We focus  on  correlators at zero frequency (finite frequency generalizations are given later).  

The zero-frequency  correlators $C_{\bf X}^{(n)}(\overline{V})$ depend on the average voltage  $\overline{V}$ over $G_1$, which is related to the voltage $V_{\rm bias}$ of the voltage source by  $\overline{V}= V_{\rm bias} (1+G_1/G_2)^{-1}$. The average voltage over $G_2$ is $V_{\rm bias} -\overline{V}= V_{\rm bias} (1+G_2/G_1)^{-1}$.  Our goal is to express  $C_{\bf X}^{(n)}(\overline{V})$ in terms of the current correlators ${\cal C}_1^{(n)}(V) $ and   ${\cal C}_2^{(n)}(V) $  that the  conductors $G_1$ and $G_2$ would have if they were isolated and biased with a non-fluctuating voltage  $V$.  These are defined by 
\begin{equation}
\langle\!\langle I_i(\omega_{1})\cdots I_i(\omega_{n})\rangle\!\rangle_{{V}}=2\pi\delta\left(\sum_{k=1}^n \omega_{k}\right){\cal C}_i^{(n)}({V}),
\end{equation}
where $I_i$ is the current through conductor $i$ at fixed voltage $V$.

To evaluate Eq.\ (\ref{eq:saddlepointF}) it is convenient to  discretize frequencies  $\omega_n = 2\pi n/\tau$. The  Fourier coefficients are   $f_n=\tau^{-1} \int_0^{\tau}{dt\, e^{i\omega_n t} \, f(t)}$. The detection time $\tau$ is sent to infinity at the end of the calculation.  For zero-frequency correlators the  sources at non-zero frequencies vanish and there is a saddle point configuration such that all fields at non-zero frequencies vanish as well.  We may then write Eq.\ (\ref{eq:saddlepointF}) in terms of only the zero-frequency fields $\vec{\Phi}_0=(V_0,\varphi_0)$,  $\vec{\Phi}_0'=(V_0',\varphi_0')$, and  $\vec{Q}_0=(I_0,q_0)$, with actions
\begin{equation}
\tau^{-1} {\cal S}_{{ k}}(\vec{{\Phi}}_0') = G_k {\varphi}_0' {V}_0'  + i \sum_{n=2}^{\infty} \frac{(-i{\varphi_0'})^n}{n!} {\cal C}_k^{(n)}({V}_0'). 
\end{equation}
For $\vec{\Phi}_0=(V_{\rm bias},0)$ and $\vec{Q}_0=(0,0)$ the saddle point is at $\vec{\Phi}_0'=(\overline{V},0)$. For the third order correlators  we need the extremum in Eq.\ (\ref{eq:saddlepointF}) to third order in $\varphi_0$ and $q_0$. We have to expand ${\cal S}_k$ to third order in the deviation  $\delta \vec{\Phi}_0'=\vec{\Phi}_0'-(\overline{V},0)$  from the saddle point at vanishing sources. We have to this order
\begin{widetext}
 \begin{eqnarray}
\tau^{-1} {\cal S}_{{ 1}}( \vec{\Phi}_0') &=&  G_1 {\varphi}_0' ( \overline{V}+ \delta {V}_0')  - { \frac{i}{2}} {\cal C}_1^{(2)}(\overline{V}) {{\varphi}}_0'^2   -  { \frac{1}{6}}  {\cal C}_1^{(3)}(\overline{V}) {{\varphi}}_0'^3-  { \frac{i}{2}} \frac{d}{d \overline{V}} {\cal C}_1^{(2)}(\overline{V}) \delta {V}_0' {{\varphi}}_0'^2  + {\cal O}(\delta \vec{\Phi}_0'^4), \\ 
\tau^{-1} {\cal S}_{{ 2}}(\vec{\Phi}_0- \vec{\Phi}_0')& =&  G_2 {\varphi}_0' (V_{\rm bias}- \overline{V}- \delta {V}_0')  - { \frac{i}{2}} {\cal C}_2^{(2)}(V_{\rm bias}-\overline{V}) {{\varphi}}_0'^2   -  { \frac{1}{6}}  {\cal C}_2^{(3)}(V_{\rm bias}-\overline{V}) {{\varphi}}_0'^3 
\nonumber \\
&&\mbox{}+ { \frac{i}{2}} \frac{d}{d \overline{V}} {\cal C}_2^{(2)}(V_{\rm bias}-\overline{V}) \delta {V}_0' {{\varphi}}_0'^2  + {\cal O}(\delta \vec{\Phi}_0'^4).
\end{eqnarray}

Minimizing the sum $ {\cal S}_{{ 1}}( \vec{\Phi}_0')+  {\cal S}_{{ 2}}(\vec{\Phi}_0- \vec{\Phi}_0')$ to third order in $q_0$ and $\varphi_0$  we then find the required relation between the correlators of the series circuit and the correlators of the isolated conductors. For the second order correlators we find
\begin{subequations}
\label{SXYresult}
\begin{eqnarray}
C^{(2)}_{II}(\overline{V})&=&(R_1+R_2)^{-2}[R_{1}^{2}{\cal C}^{(2)}_{1}(\overline{V})+R_{2}^{2}{\cal C}^{(2)}_{2}(V_{\rm bias}-\overline{V})],\label{SIIresult}\\
C^{(2)}_{VV}(\overline{V})&=&(R_1+R_2)^{-2}(R_{1}R_{2})^{2}[{\cal C}^{(2)}_{1}(\overline{V})+{\cal C}^{(2)}_{2}(V_{\rm bias}-\overline{V})],\label{SVVresult}\\
C^{(2)}_{IV}(\overline{V})&=&(R_1+R_2)^{-2}R_{1}R_{2}[R_{2}{\cal C}^{(2)}_{2}(V_{\rm bias}-\overline{V})-R_{1}{\cal C}^{(2)}_{1}(\overline{V})],\label{SIVresult}
\end{eqnarray}
\end{subequations}
with $R_k=1/G_k$. The third order correlators contain extra terms that depend on the second-order correlators,
\begin{subequations}
\label{eq:Cresult}
\begin{eqnarray}
C_{III}^{(3)}(\overline{V})&=&(R_1+R_2)^{-3}[R_{1}^{3}{\cal C}_{1}^{(3)}(\overline{V})+R_{2}^{3}{\cal
C}_{2}^{(3)}(V_{\rm bias}-\overline{V})]+3 C_{IV}^{(2)} \frac{d}{d\overline{V}}C_{II}^{(2)},
\label{CIresult}\\
C_{VVV}^{(3)}(\overline{V})&=&(R_1+R_2)^{-3}(R_{1}R_{2})^{3}[{\cal C}_{2}^{(3)}(V_{\rm bias}-\overline{V})-{\cal
C}_{1}^{(3)}(\overline{V})]+3C^{(2)}_{VV}
\frac{d}{d\overline{V}}C^{(2)}_{VV},\label{CVresult}\\
C_{VVI}^{(3)}(\overline{V})&=&(R_1+R_2)^{-3}(R_{1}R_{2})^{2}[R_{1}{\cal
C}_{1}^{(3)}(\overline{V})+R_{2}{\cal C}_{2}^{(3)}(V_{\rm bias}-\overline{V})]+2C^{(2)}_{VV}
\frac{d}{d\overline{V}}C^{(2)}_{IV}+C^{(2)}_{IV}
\frac{d}{d\overline{V}}C^{(2)}_{VV},\label{CVVIresult}\\
C_{IIV}^{(3)}(\overline{V})&=&(R_1+R_2)^{-3}R_{1}R_{2}[R_{2}^{2}{\cal
C}_{2}^{(3)}(V_{\rm  bias}-\overline{V})-R_{1}^{2}{\cal C}_{1}^{(3)}(\overline{V})]+
2C^{(2)}_{IV}\frac{d}{d\overline{V}}C^{(2)}_{IV}+C^{(2)}_{VV}
\frac{d}{d\overline{V}}C^{(2)}_{II}.\label{CIIVresult}
\end{eqnarray}
\end{subequations}
\end{widetext}
These results agree with those obtained by the cascaded Langevin approach \cite{Bee03}. 

 \subsection{Mesoscopic and macroscopic conductor in series}
 
 An important application is a single mesoscopic conductor $G_1$  embedded in an electromagnetic environment, represented by a macroscopic conductor $G_2$. A macroscopic conductor has no shot noise but only thermal noise. The third cumulant ${\cal C}_2^{(3)}$ is therefore equal to zero. The second cumulant ${\cal C}_2^{(2)}$ is voltage independent, given by    \cite{Bla00}
\begin{equation} \label{eq:envnoise}
{\cal C}^{(2)}_{2}(\omega)= \omega \coth\left(\frac{\omega}{2kT_2}\right) {\rm  Re} \,G_2(\omega),
\end{equation}
at temperature $T_2$.  We still assume low frequencies $\omega \ll \max(e\overline{V},kT_1)$, so the frequency dependence of ${\cal S}_1$ can be neglected. We have retained the frequency dependence of ${\cal S}_2$, because the characteristic frequency of a macroscopic conductor is typically much smaller than of a mesoscopic conductor. 

From   Eq.\ (\ref{eq:Cresult})  (and a straightforward generalization to frequency dependent correlators) we can obtain the third cumulant correlators by setting ${\cal C}_2^{(3)}=0$ and substituting Eq.\ (\ref{eq:envnoise}).  We only give the two correlators $ C_{III}^{(3)}$ and $C_{VVV}^{(3)}$,  since these are the most significant for experiments. To abbreviate the formula we denote $G=G_1$ and $Z(\omega)=1/G_2(\omega)$.  We find  
\begin{widetext}
\begin{eqnarray} \label{eq:C3I}
 C_{III}^{(3)}(\omega_{1},\omega_{2},\omega_{3})&=&\frac{{\cal C}_1^{(3)}(\overline{V})-(d{\cal C}_1^{(2)}/d\overline{V})\sum_{j=1}^3 Z(-\omega_{j})[{\cal  C}_1^{(2)}(\overline{V})-G Z(\omega_j) {\cal C}^{(2)}_{2}(\omega_{j})][1+Z(-\omega_{j})G]^{-1}}{[1+Z(\omega_{1})G][1+Z(\omega_{2})G][1+Z(\omega_{3})G]}, \\
-\frac{C_{VVV}^{(3)}(\omega_{1},\omega_{2},\omega_{3})}{Z(\omega_{1})Z(\omega_{2})Z(\omega_{3})}&=& \frac{{\cal C}_1^{(3)}(\overline{V})-(d{\cal C}_1^{(2)}/d\overline{V})\sum_{j=1}^3Z(-\omega_{j})[{\cal C}_1^{(2)}(\overline{V})+ {\cal C}^{(2)}_{2}(\omega_{j})][1+Z(-\omega_{j})G]^{-1}}{[1+Z(\omega_{1})G][1+Z(\omega_{2})G][1+Z(\omega_{3})G]}.\label{eq:C3V}
\end{eqnarray} 
\end{widetext} 

We show plots for two types of mesoscopic conductors: a tunnel junction and a diffusive metal. In both cases it is assumed that there is no inelastic scattering, which is what makes the conductor mesoscopic. 
 The plots correspond to global thermal equilibrium ($T_1=T_2=T$) and to a real and frequency-independent impedance $Z(\omega)\equiv Z$. We compare $C_I^{(3)}\equiv C_{III}^{(3)}$ with  $C_V^{(3)}\equiv -C_{VVV}^{(3)}/Z^3$. (The minus sign is chosen so that $C_I^{(3)}=C_V^{(3)}$ at $T=0$.)
 
\begin{figure}
\includegraphics[width=12cm]{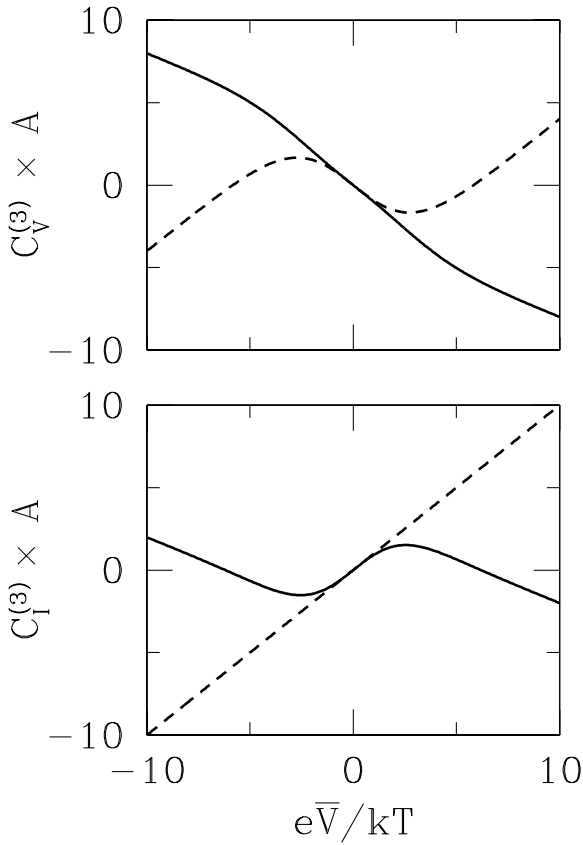}
\caption{
Third cumulant of voltage and current fluctuations of a tunnel junction (conductance $G$) in an  electromagnetic environment (impedance $Z$, assumed frequency independent). Both $C^{(3)}_I$ and $C^{(3)}_V$ are multiplied by the scaling factor $A=(1+ZG)^3/eGkT$.    The two curves correspond to different values of $ZG$ (solid curve: $ZG=1$; dashed curve: $ZG=0$).   The temperatures of the tunnel junction and its environment are chosen the same, $T_1=T_2=T$. \label{C3exp}
 }
\end{figure}

 For a tunnel junction one has 
\begin{equation} \label{eq:tunnelnoise}
  {\cal C}_1^{(2)}(V)= G eV  \coth\frac{e V}{2 k  T_{ }}, \;\;\;
 {\cal C}_1^{(3)}(V)=G e^2  V  .
\end{equation}
The third cumulant of current fluctuations in an isolated tunnel junction  is temperature independent \cite{Lev01}, but this is changed drastically by the electromagnetic environment \cite{Bee03}. Substitution of Eq.\ (\ref{eq:tunnelnoise}) into Eqs.\ (\ref{eq:C3I}) and (\ref{eq:C3V}) gives the curves plotted in Fig.\ \ref{C3exp}  for $ZG=0$ and $ZG=1$. The slope $dC^{(3)}_V(\overline{V})/d\overline{V}$ becomes strongly temperature dependent and may even change sign when $kT$ becomes larger than $e\overline{V}$. This is in qualitative agreement with the experiment of Reulet, Senzier, and Prober \cite{Reu03}. In Ref.\ \cite{Reu03} it is shown that Eq.\ (\ref{eq:C3V}) provides a quantitative description of the experimental data.  

\begin{figure}
\includegraphics[width=12cm]{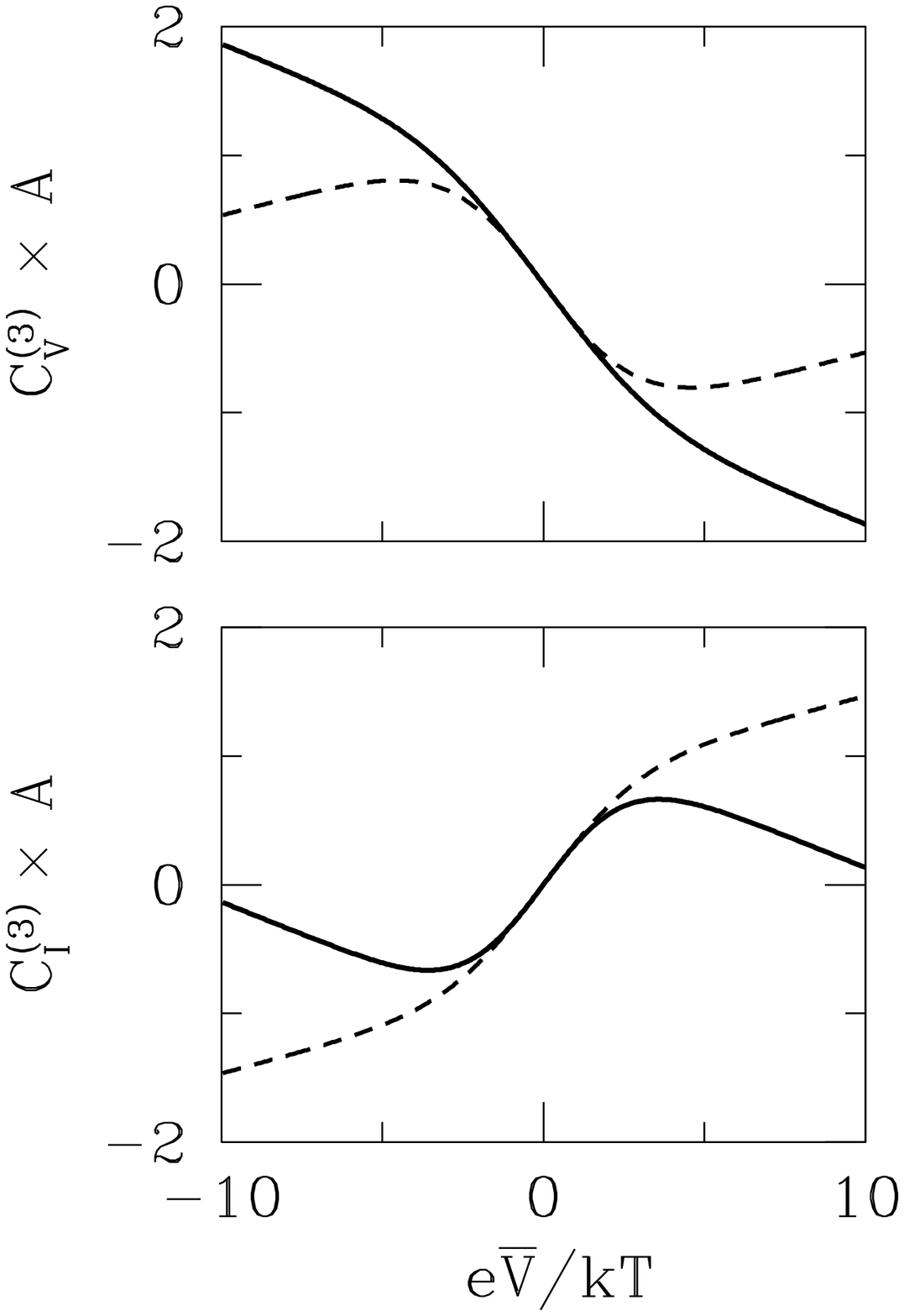}
\caption{
Same as Fig.\ \ref{C3exp}, but now for a diffusive metal.  \label{diffusive}
 }
\end{figure}

 For a diffusive metal we substitute the known formulas for the second and third cumulants without electromagnetic environment  \cite{Gut02,Nag02},
  \begin{eqnarray}
 {\cal C}_1^{(2)}(V)& =& \frac{1}{3} GeV\left( \coth p + 2/p\right), \\
 {\cal C}_1^{(3)}(V) &=& e^2 GV \frac{p(1-26 e^{2p}+e^{4p}) - 6(e^{4p}-1)}{15p(e^{2p}-1)^2}. \nonumber \\
\end{eqnarray}
We have abbreviated $p=e{V}/2 k T$.
Plots for $ZG=0$ and $ZG=1$  are shown in Fig.\ \ref{diffusive}. The diffusive metal is a bit less striking than  a tunnel junction, since the third
cumulant is already temperature dependent even in the absence of the electromagnetic environment. In the limit $ZG \to \infty$ we recover the result for $C^{(3)}_V$  obtained by Nagaev from the cascaded Langevin approach \cite{Nag03}.

\section{How to measure current fluctuations} \label{sc:current}

In Fig.\ \ref{C3exp}  we have plotted both current and voltage correlators, but only the voltage correlator has been measured \cite{Reu03}. At zero temperature of the macroscopic conductor there is no difference between the two, as follows from Eqs.\ (\ref{eq:C3I}) and (\ref{eq:C3V}): $C_{III}^{(3)} =- C_{VVV}^{(3)}/Z^3$ if ${\cal C}_2^{(2)}=0$, which is the case for a macroscopic conductor $G_2$ at $T_2=0$. For $T_2\neq 0$ a difference appears that persists in the limit of a non-invasive measurement $Z\to 0$ \cite{Bee03}.  Since $V$ and $I$  in the series circuit with a macroscopic $G_2$ are linearly related and   linear systems are known to be  completely determined by their response functions and their temperature, one could ask what it is that  distinguishes the two measurements, or more practically: How would one measure $C_{III}^{(3)}$ instead of $C_{VVV}^{(3)}$?  

 \begin{figure}
\includegraphics[width=3cm,angle=270]{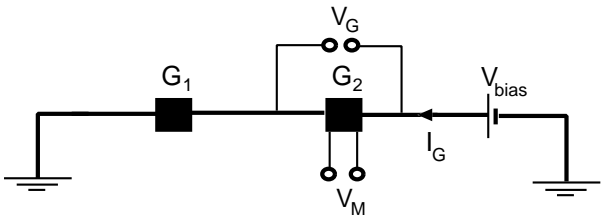}
\caption{Four-terminal voltage measurement.   }
\label{fourterm}
\end{figure}
 
 To answer this question we slightly generalize the macroscopic conductor to a four-terminal, rather than two-terminal configuration, see Fig.\ \ref{fourterm}. The voltage $V_M$ over the extra pair of contacts is related to the current $I_G$ through the series circuit by a cross impedance, $\partial V_M/\partial I_G = Z_{MG}$. The full impedance matrix ${\bf Z}$ is defined as in Eq.\ (\ref{eq:Zmatrix}).  For simplicity we  take the zero-frequency limit.  For this configuration the third cumulant $C^{(3)}_{V_MV_MV_M}$  of $V_M$ is given by 
 \begin{eqnarray} \label{eq:interpolate}
  \frac{C^{(3)}_{V_MV_MV_M}}{Z^3_{MG}}&=&  C^{(3)}_{I_GI_GI_G} \nonumber \\
&&\!\!\!\!\!\!\!\!\!\! \!\!\!\!\!\!\!\!\!\! \!\!\!\!\!\!\!\!\!\! \mbox{} +\frac{Z_{GM}+Z_{MG}}{2 Z_{GM}}\left( \frac{ C^{(3)}_{V_GV_GV_G}}{Z^3_{GG}} -C^{(3)}_{I_GI_GI_G}\right).
 \end{eqnarray}
 It contains  the correlator  $\langle \! \langle \delta V_M(\omega) \delta V_G(\omega') \rangle \! \rangle = 2\pi \delta(\omega+\omega') C_{GM}$ of the voltage fluctuations over the two pairs of terminals of the macroscopic conductor, which according to the fluctuation-dissipation theorem (\ref{eq:Gaussnoise}) is given in the zero-frequency limit  by 
 \begin{equation}
 C_{GM} = kT_2 (Z_{GM} + Z_{MG}).
 \end{equation}
 The correlator $C_{GM}$ enters since  $C^{(3)}_{V_MV_MV_M}$ depends on how thermal fluctuations in the measured variable $V_M$ correlate with the thermal fluctuations of $V_G$ which  induce extra current noise in $G_1$.
 
 We conclude from Eq.\ (\ref{eq:interpolate}) that the voltage correlator  $C^{(3)}_{V_MV_MV_M}$ becomes proportional to the current correlator $ C^{(3)}_{I_GI_GI_G} $ if $Z_{GM} + Z_{MG}=0$. 
 This can be realized  if $V_M$ is the Hall voltage $V_H$  in a weak magnetic field $B$. Then $Z_{MG}=-Z_{GM}=R_H$, with $R_H\propto |B|$ the Hall resistance.  The magnetic field need only be present in the macroscopic conductor $G_2$, so it need not disturb the transport properties of the mesoscopic conductor $G_1$.  If, on the other hand, $V_M$ is the longitudinal voltage $V_L$, then $Z_{MG}=Z_{GM}=R_L$, with $R_L$ the longitudinal resistance.  The two-terminal impedance $Z_{GG}$ is the sum of  Hall and longitudinal resistances, $Z_{GG}=R_L+R_H$. So one has   
 \begin{eqnarray}
\  C_{V_LV_LV_L }^{(3)}&=&  \left(\frac{R_L}{R_L+R_H}\right)^3 C_{V_GV_GV_G }^{(3)}, \\
 C_{V_HV_HV_H }^{(3)} &=&  R_H^3 C_{I_GI_GI_G}^{(3)}.
\end{eqnarray}
  \begin{figure}
\includegraphics[width=8cm]{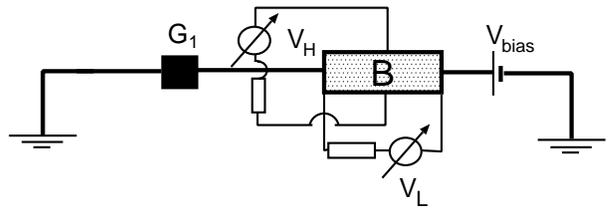}
\caption{Hall bar that allows one to measure the voltage correlator  $C_{V }^{(3)}\propto\langle \!\langle V_L^3 \rangle\! \rangle$  as well as the current correlator $C_{I }^{(3)}\propto\langle \!\langle V_H^3 \rangle\! \rangle$.  }
\label{fighall}
\end{figure}

One can  generalize all this to an arbitrary measurement variable $X$ that is linearly related to the current $I_G$ through $G_1$.  In a linear circuit the off-diagonal elements of the response tensor ${\bf Z}$ relating   $(X,V_{G})$ to the conjugated sources are linked by Onsager-Casimir  relations \cite{Cas45}. If $X$ is even under time-reversal, then $Z_{XG}=Z_{GX}$, while if $X$ is odd, then $Z_{XG}=-Z_{GX}$.  In the first case  $ C^{(3)}_{XXX}\propto C^{(3)}_{V_GV_GV_G}$, while in the second case $ C^{(3)}_{XXX}\propto C^{(3)}_{I_GI_GI_G}$.

\section{Environmental Coulomb blockade} \label{sc:Coulomb}

  The saddle-point approximation to the  path integral  (\ref{eq:convolution}) for a mesoscopic conductor $G_1$ in series with a macroscopic conductor $G_2$ (impedance $Z$) breaks down when the impedance at the characteristic frequency scale  $\Lambda= 1/\max(\tau_1,\tau_2)$ discussed in section \ref{sc:linres}  is not small compared to the resistance quantum $h/e^2$.   It can then react fast enough to affect the dynamics of the  transfer of a single electron.  These single-electron effects amount to a Coulomb blockade induced by the electromagnetic environment \cite{IngNaz}.   In our formalism they are accounted for by fluctuations around the saddle point of Eq.\ (\ref{eq:convolution}). 
 
 In Ref.\  \cite{Yey01} it has been found that the Coulomb blockade correction to the mean current calculated to leading order in  $Z$  is proportional to the second cumulant of current fluctuations in the isolated mesoscopic conductor ($Z=0$). More recently, the  Coulomb blockade correction to the second cumulant of current fluctuations has been found to be proportional to the third cumulant \cite{Gal02}.  It was conjectured in Ref.\ \cite{Gal02} that this relation holds also for higher cumulants. Here we give a proof of this conjecture. 
 
We show that at zero temperature and zero frequency the leading order Coulomb blockade  correction to the $n$-th cumulant of current fluctuations  is proportional to the voltage derivative of the $(n+1)$-th cumulant.  
To extract  the environmental Coulomb blockade from the other effects of the environment  we assume that $Z$ vanishes at zero frequency, $Z(0)=0$.  The derivation is easiest in terms of the pseudo-probabilities discussed in Sec.\ \ref{sc:calculation}. 

According to Eq.\ (\ref{eq:currint}),
 cumulant correlators of current have the generating functional 
 \begin{equation}
{\cal F}_{ G_1+G_2}[\vec{\Phi}_{ }=(V,\varphi)] = \ln \int{{\cal D} I \, e^{-i\int{dt\, I \varphi }}\, {\cal P}_{ G_1+G_2}[V,I] }.
\end{equation}
Zero frequency current correlators are obtained from 
\begin{equation} \label{eq:cumcorr}
\langle \!\langle I(0)^n \rangle\! \rangle_{ G_1+G_2} = i^n\frac{\delta^n}{\delta [\varphi(0)]^n} {\cal F}_{ G_1+G_2}[\vec{\Phi}_{ }]\Big|_{\varphi=0}.
\end{equation}
We employ now Eq.\ (\ref{eq:pseudorellin}) and  expand  ${\cal F}_{ G_1+G_2}[\vec{\Phi}_{ }]$ to first order in $Z$,
\begin{eqnarray}
&&{\cal F}_{ G_1+G_2}[\vec{\Phi}_{ }]= {\cal F}_{G_1}[\vec{\Phi}_{ }] \nonumber \\
&&\;\;\;\mbox{}-\frac{ \int{{\cal D}I\,e^{-i\int{dt\, I \varphi }}\int{\frac{d\omega}{2\pi} \, Z(\omega) I(\omega) \frac{\delta}{\delta V(\omega)} {\cal P}_{ G_1}[V,I]}}}{ \int{{\cal D}I\, e^{-i\int{dt\, I \varphi }}}\,  {\cal P}_{ G_1}[V,I]} \nonumber \\
 \;\;\;&&\mbox{}={\cal F}_{G_1}[\vec{\Phi}_{ }]- i\int{\frac{d\omega}{2\pi} \, Z(\omega) \frac{\delta^2}{\delta V(\omega)\delta \varphi(\omega)} {\cal F}_{G_1}[\vec{\Phi}_{ }]}.\nonumber \\
\end{eqnarray}
The last equality holds since single derivatives of ${\cal F}_{G_1}[\vec{\Phi}]$ with respect to a variable at finite frequency vanish because of time-translation symmetry.   Substitution into Eq.\ (\ref{eq:cumcorr}) gives
\begin{eqnarray}
\langle \!\langle I(0)^n \rangle\! \rangle_{ G_1+G_2}& =&  \langle \!\langle I(0)^n \rangle\! \rangle_{ G_1} \nonumber \\
&&\!\!\! -\int{\frac{d\omega}{2\pi} \, Z(\omega) \frac{\delta}{\delta V(\omega)} \langle \!\langle I(\omega) I(0)^n \rangle\! \rangle_{ G_1}},   \nonumber \\
\end{eqnarray}
which is what we had set out to prove.

\section{Conclusion} \label{sc:conclusion}

In conclusion, we have presented a fully quantum mechanical 
derivation of the effect of an electromagnetic environment on current 
and voltage fluctuations in a mesoscopic conductor, going beyond an 
earlier study at zero temperature \cite{Kin02}. The results agree with 
those obtained from the cascaded Langevin approach \cite{Bee03}, 
thereby providing the required microscopic justification.

 From an experimental point of view, the nonlinear feedback from the 
environment is an obstacle that stands in the way of a measurement of 
the transport properties of the mesoscopic system. To remove the 
feedback it is not sufficient to reduce the impedance of the 
environment. One also needs to eliminate the mixing in of 
environmental thermal fluctuations. This can be done by ensuring that 
the environment is at a lower temperature than the conductor, but 
this might not be a viable approach for low-temperature measurements. 
We have proposed here an alternative method, which is to ensure that 
the measured variable changes sign under time reversal. In practice 
this could be realized by measuring the Hall voltage over a 
macroscopic conductor in series with the mesoscopic system.

The field theory developed here also provides for a systematic way to 
incorporate the effects of the Coulomb blockade which arise if the 
high-frequency impedance of the environment is not small compared to 
the resistance quantum. We have demonstrated this by generalizing to 
moments of arbitrary order a relation in the literature 
\cite{Yey01,Gal02} for the leading-order Coulomb blockade correction to 
the first and second moment of the current. We refer to Ref.\ 
\cite{Kin03} for a renormalization group analysis of the Coulomb 
blockade corrections of higher order.

\begin{acknowledgements}
We  thank D. Prober and B. Reulet for discussions of their experiment. This research was supported by the ``Ne\-der\-land\-se
or\-ga\-ni\-sa\-tie voor We\-ten\-schap\-pe\-lijk On\-der\-zoek'' (NWO)
and by the ``Stich\-ting voor Fun\-da\-men\-teel On\-der\-zoek der
Ma\-te\-rie'' (FOM).
\end{acknowledgements}

\begin{appendix}

\section{Derivation of the environmental action} \label{app:action}

To derive Eq.\ (\ref{eq:envaction}) 
we define a generating functional for the voltages ${\bf V}=(V_M,V_G)$ in the environmental  circuit of Fig.\ \ref{linear},
\begin{equation} \label{eq:Zlinear}
 {\cal Z}_{ e} [ \boldsymbol{ \vec{Q}}]= \left\langle  
 { T}_- e^{i \int{dt\; \left[{H}_{} +\boldsymbol{  Q^-}(t)  \boldsymbol{ V}\right]} }  { T}_+ e^{-i \int{dt\; \left[{H} + \boldsymbol{ Q^+}(t)  \boldsymbol{ V}\right]} } \right\rangle.
\end{equation}
We have introduced sources  $\boldsymbol{Q}=({Q_M},{Q_G})$. Since the environmental Hamiltonian is quadratic, the    generating functional  is the exponential of a quadratic form in   $\boldsymbol{\vec{Q}}$,  \begin{equation} \label{eq:Zj}
 {\cal Z}_{ e}[\boldsymbol{\vec{Q}}] = \exp\left( - { \frac{i}{2}} \int{ \frac{d \omega}{2 \pi}\, \boldsymbol{\vec{Q}}^{\dagger} (\omega)\boldsymbol{\check{G}}(\omega) \boldsymbol{\vec{Q}}(\omega)} \right).
\end{equation}
  The off-diagonal elements of the  matrix $\boldsymbol{\check{G}}$ are determined by the impedance of the circuit,  
\begin{eqnarray} \label{eq:responseZ}
 &&i  \frac{\delta^2}{ \delta Q_{\beta}^{cl}(\omega') \delta Q^{q*}_{\alpha}(\omega)}\ln {\cal Z}_{ e} \Big|_{\boldsymbol{\vec{Q}}=0}=   \frac{\delta}{\delta I_{\beta}(\omega')} \langle V_{\alpha}(\omega) \rangle \nonumber \\
 && =2\pi \delta(\omega-\omega') Z_{\alpha\beta}(\omega) 
 .
\end{eqnarray}
The upper-diagonal ($cl,cl$) elements in the Keldysh space vanish for symmetry reasons (${\cal Z}_e|_{Q^q=0}=0$, cf.\ Ref.\ \cite{Kam01}). The lower-diagonal ($q,q$)  elements are determined by the fluctuation-dissipation theorem (\ref{eq:Gaussnoise}),
 \begin{eqnarray} \label{eq:flucdiss}
 &&-  \frac{\delta^2}{ \delta Q_{\alpha}^{q*}(\omega) \delta Q^{q*}_{\beta}(\omega')}\ln {\cal Z}_{ e} \Big|_{\boldsymbol{\vec{Q}}=0} =  \langle \delta V_{\alpha}(\omega) \delta V_{\beta}(\omega') \rangle \nonumber \\
 &&=\pi  \delta(\omega+\omega')  \omega \coth\left(\frac{\omega}{2k T}\right) 
   \, [Z_{\alpha \beta}(\omega)+Z^*_{\beta\alpha}(\omega)] . \nonumber \\
   \end{eqnarray}
  Consequently we have
  \begin{equation} \label{eq:G}
\boldsymbol{\check{G}}(\omega) =  \left( \begin{array}{cc}
0 &  \boldsymbol{ Z}_{}^{\dagger}(\omega) \\
  \boldsymbol{ {Z}} (\omega) & -\frac{i}{2}\omega  \coth\left(\frac{ \omega}{2k T}\right)[\boldsymbol{ {Z}} (\omega)+ \boldsymbol{ Z}^{\dagger}(\omega)]
\end{array} \right). 
\end{equation}
The environmental action ${\cal S}_e$  is defined by 
 \begin{equation} \label{eq:envint} 
 {\cal Z}_{ e}[\boldsymbol{ \vec{Q}}]=  \int{ {\cal D}[\boldsymbol{\vec{\Phi}_{ G}}] \exp \left(-i {\cal S}_{ e}[\boldsymbol{ \vec{Q}_{ M}},\boldsymbol{\vec{\Phi}_{ G}}]-i \boldsymbol{\vec{\Phi}_G} \times \boldsymbol{\vec{Q}_G}    \right)}. 
 \end{equation}
 One can check that substitution of Eq.\ (\ref{eq:envaction}) into Eq.\ (\ref{eq:envint}) yields the same ${\cal Z}_e$ as given by Eqs.\ (\ref{eq:Zj}) and (\ref{eq:G}).

\section{Derivation of Eq.\ (3.14) } \label{app:ZIZV}

\begin{figure}
\includegraphics[width=4cm]{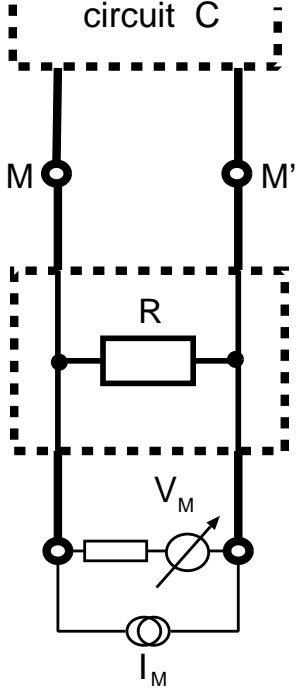}
\caption{Circuit to relate voltage to current measurements.  }
\label{figapp}
\end{figure}

 In the limit $R\to \infty$ a voltage measurement in the circuit of Fig.\ \ref{figapp} corresponds to a voltage measurement at contacts $M$ and $M'$ of the  circuit $C$. We obtain the generating functional ${\cal Z}_V$ of this voltage measurement from  Eq.\ (\ref{eq:result}). The influence functional is now due to $C$ and it equals the generating functional ${\cal Z}_I$  of a current measurement at contacts $M$ and $M'$ of $C$.   From Eq.\ (\ref{eq:envaction})  with $Z_{MM}=Z_{GG}=-Z_{MG}=-Z_{GM}=R$ we find  in the limit  $R\to \infty$ that the environmental action takes the simple form  ${\cal S}_e[\vec{Q}_M,\vec{\Phi}_G]= \vec{\Phi} \times \vec{Q}$, with the cross-product defined in Eq.\ (\ref{eq:cross}).  Consequently, we have   
\begin{equation}  
{\cal Z}_{V}[\vec{Q}]=   \int{  {\cal D}[\vec{\Phi}] \, e^{-i\vec{\Phi}\times \vec{Q} }\,
{\cal Z}_{I}[\vec{\Phi}]} .
\end{equation}
This equation relates the generating functionals of current and voltage measurements at any pair of contacts of a circuit. 

\section{Derivation of Eq.\ (4.1)}
\label{app:8term}

To derive Eq.\ (\ref{eq:convolution}) from Eq.\ (\ref{eq:result}) we need the environmental action ${\cal S}_e$ of the circuit shown in  Fig.\ \ref{eightterminal}.  The impedance matrix is
\begin{widetext}
\begin{equation}
\boldsymbol{Z} = \frac{1}{R_1+R_2+R_3}
\left( \begin{array}{cccc}
R_1(R_2+R_3) & - R_1 R_2 & -R_1(R_2+R_3) & - R_1 R_3   \\  
-R_1 R_2 & R_2(R_1+R_3) & -R_1 R_2 & - R_2 R_3 \\
-R_1(R_2+R_3) & -R_1 R_2 & R_1(R_2+R_3) &  -R_1 R_3 \\
-R_1 R_3 & - R_2 R_3  & -R_1 R_3 & R_3(R_1+R_2) 
\end{array} \right).
\end{equation}
\end{widetext}
We seek the limit  $R_1,R_2,R_3 \to \infty$.  The environmental action  (\ref{eq:envaction}) takes   the form
\begin{equation}
{\cal S}_{ e}[\boldsymbol{ \vec{Q}_{ M}},\boldsymbol{\vec{\Phi}_{ G}}]=
 \vec{\Phi}_{G_1} \times \vec{Q}_{M_1} +\vec{\Phi}_{G_1} \times \vec{Q}_{M_2} +\vec{\Phi}_{G_2} \times \vec{Q}_{M_2} .
\end{equation}
Substitution into Eq.\ (\ref{eq:result}) gives  ${\cal Z}_{VV}$.
Employing Eq.\ (\ref{eq:ZIZV}) to obtain ${\cal Z}_{VI}$ from ${\cal Z}_{VV}$  we  arrive at  Eq.\ (\ref{eq:convolution}).

\end{appendix}


\begin{thebibliography}{99}
\bibitem{Ben83} E. Ben-Jacob, E. Mottola, and G. Sch\"{o}n, Phys.\ Rev.\ Lett.\
{\bf  51}, 2064 (1983).
\bibitem{Sch85}
 G. Sch\"{o}n, Phys.\ Rev.\ B {\bf  32}, 4469 (1985).
\bibitem{Dev90} M. H. Devoret, D. Esteve, H. Grabert, G.-L. Ingold, H. Pothier, and C. Urbina, Phys.\ Rev.\ Lett. {\bf 64}, 1824 (1990).
\bibitem{IngNaz} G.-L. Ingold and Yu.\ V. Nazarov, in {\em Single Charge
Tunneling}, edited by H. Grabert and M. H. Devoret, NATO ASI Series B294
(Plenum, New York, 1992).
 \bibitem{Lee96} H. Lee and L. S. Levitov, Phys.\ Rev.\ B {\bf  53}, 7383 (1996).
\bibitem{Shulman} Sh. Kogan, {\it Electronic Noise and Fluctuations in Solids} (Cambridge  University, Cambridge, 1996).
\bibitem{Bla00} Ya.\ M. Blanter and M. B\"{u}ttiker, Phys.\ Rep.\ {\bf  336}, 1
(2000).  
\bibitem{Bee02}
C. W. J. Beenakker and C. Sch\"onenberger, Physics Today {\bf 56} (5), 37 (2003).
\bibitem{Kin02} M. Kindermann, Yu.\ V. Nazarov, and C. W. J. Beenakker, Phys. Rev. Lett. (in press).
\bibitem{Bee03} C. W. J. Beenakker, M. Kindermann, and Yu. V. Nazarov, Phys. Rev. Lett. {\bf 90}, 176802 (2003). 
\bibitem{Lev01} L. S. Levitov and M. Reznikov, cond-mat/0111057.
\bibitem{Gut02} D. B. Gutman and Y. Gefen, cond-mat/0201007.
\bibitem{Nag02} K. E. Nagaev, Phys.\ Rev.\ B {\bf  66}, 075334 (2002).
\bibitem{Reu03} B. Reulet, J. Senzier, and D. E. Prober, cond-mat/0302084; B. Reulet, private communication. 
\bibitem{Naz99} Yu.\ V. Nazarov, Ann.\ Phys.\ (Leipzig) {\bf  8}, 507 (1999).
\bibitem{Kin01} Yu.\ V. Nazarov and M. Kindermann, cond-mat/0107133.
\bibitem{Yey01}
A. Levy Yeyati, A. Martin-Rodero, D. Esteve, and C. Urbina, Phys. Rev. Lett. {\bf  87}, 046802  (2001).
\bibitem{Gal02} A. V. Galaktionov, D. S. Golubev, and A.  D.  Zaikin, cond-mat/0212494.
\bibitem{Kin03}
M. Kindermann and Yu. V. Nazarov, cond-mat/0304078.
\bibitem{Cas45}   H. B. G. Casimir, Rev. Mod. Phys. {\bf  17}, 343 (1945).
\bibitem{Kam01} A. Kamenev,  in {\it Strongly Correlated Fermions and Bosons in Low-Dimensional Disordered Systems}, edited by I. V. Lerner, B. L. Altshuler, V. I. Fal'ko, and T. Giamarchi, NATO Science Series II Vol.\ 72 (Kluwer, Dordrecht, 2002); cond-mat/0109316.
\bibitem{review} M. Kindermann and  Yu. V. Nazarov, in {\it Quantum Noise}, edited by Yu. V. Nazarov and Ya.\ M. Blanter, NATO Science Series II Vol.\ 97 (Kluwer, Dordrecht, 2003); cond-mat/0303590.
\bibitem{Bel01} W. Belzig and Yu. V. Nazarov, Phys. Rev. Lett. {\bf 87}, 197006 (2001).
\bibitem{Gav01}
U. Gavish, Y. Imry, L. Levinson, and B. Yurke,  in {\it Quantum Noise}, edited by Yu. V. Nazarov and Ya.\ M. Blanter, NATO Science Series II Vol.\  97  (Kluwer, Dordrecht, 2003);  cond-mat/0211646.
\bibitem{Lev93} L. S. Levitov and G. B. Lesovik, JETP Lett.\ {\bf  58}, 230
(1993).
\bibitem{Lev96}
 L. S. Levitov, H. Lee, and G. B. Lesovik, J. Math.\
Phys.\ {\bf  37}, 4845 (1996).
\bibitem{Nag03} K. E. Nagaev, cond-mat/0302008.


\end{thebibliography}
\end{document}